# SMART GRID DEMAND MONITORING MODEL

Ms.Kalpana Kandpal
Dept of Computer Science
Mahamaya Technical University

Ms.Anjali Singhal
Dept of Computer Science
Mahamaya Technical Unversity

*Abstract*— **Smart Grid (SG) is a smart concept based on cerebral advanced technologies; SG also becomes a subject matter for research. It unified inventive machine and technologies starts from generation, transmission and distribution to the consumer usage. With the advancement in the technology, various steps are taking place to convert power grid to smart grid.**

**At current stage various models of SG are came in existence like Smart Grid Investment Model, Smart Grid Interoperability Maturity Model, Smart Grid Conceptual Model and Smart Grid Monitoring Model are in hand. Smart Grid Interoperability Maturity Model checks or automate the status of power at its various sites, Smart Grid Investment Model is used to estimate various investment done in grid, Smart grid Conceptual Model narrate about various paradigm and ability, Smart Grid Monitoring Model narrate about arrangement and proficiency within electric service areas. A new proposed Smart Grid Demand Monitoring Model will make us able to monitor the demand management.**

## I. INTRODUCTION

The power grid provides power from generation to distribution and next to consumer (domestic, commercial). The procedure have basically two system i.e. Transmission system (TS) and Distribution System (DS). The (TS) provide power from power plant to distribution system and it provide next to consumers. Power grid is an interconnected transmission system using analog technology.

Recent technology of Smart Grid is spreading all around that is a tenuous term on both sides with different functionality jumping to remodel the power grid. At its center, a smart grid provides digital communication and control system to observe and regulate the power flows.

Smart Grid updates the coordination, connection and automation between supplier, consumer and network by advanced grid feature as demand management, generation, real-time pricing.

Till date various models are present for monitoring and measuring an organization progress to many considerable correspondents like reliability, efficiency, automation, integration and its result on economy.

## II. AVAILABLE DEMAND APPROACHES AND SOFTWARE MODELS

### A. An Informatics Approach to Demand Response Optimization in Smart Grids:

The core of this report is to retrieve a methodology which will helps in data management and to construct smart grid tool that cerebrally attains the growing necessity for power and report by improving power utilization in the city. This provides a accessible and protected software design for demand response utilization, that can alter the self-motivated environment and regulates the difficulty produced by date torrent from Advanced Metering Infrastructure and other sources.

### B. Autonomous Demand-Side Management Based on Game-Theoretic Energy Consumption Scheduling for the Future Smart Grid:

The majority of present demand-side management program cores on the interrelationship between an effective company and its consumer. This research is done to present a separated and scattered demand-side power controlling system among the consumers who takes the benefit of mutual digital communication architecture which can be proposed in further smart grid.

Here a game theory is used to create an energy consumption program, where the consumer will be the player who has to daily looked after their whole house hold equipment and their loads. Now it is supposed that the electric company can use an effective pricing tariff that narrate the consumption holding various consumers, the universal effective work done by reducing the cost of energy acquired at equilibrium of converted energy consumption planning game. The consumers proceed with secrecy and will not show their consumption report to other consumer.

*C. Smart Grid Interoperability Maturity Model (SGIMM):*

SGIMM was settled by the Grid Wise Architecture Council (GWAC) to observe and calculate the computation in the area transmission, distribution and demand side resources. The characteristics are as follows:

- Prestige/advancement calculating figures.
- Interruption inquiry.
- Precedence of strength to enhance the existing prestige.

The key feature of SGIMM is to develop a measuring model that can provide interaction of main subject like alignment and progression, procedure and performance, safety and protection in utility control scheme. To achieve this edge among various result and scheme with clear objective and principle are defined.

These objectives are then transformed into a sequence of measures. By the help of these measures, the presentation and advancement results and calculated. If transformation is countered, then the new principle and the set of goals are defined. SGIMM helps in improvement of edge arrangement involved in smart grid technology.

*D. Smart Grid Monitoring Model(SGMoM):*

This model was designed for the appraisal of the some of the component of smart grid like people, process and technology. It also helps in the evaluation of the framework and communication principle for monitoring and management technologies so that a proper assumption about a thing is organized. Task performed are as follow:

- Solemnize recent understanding of system procedure and changing aspects.
- Narrate about the connection of procedure through corrective edge.
- Narrate the assured opportunity of the procedure that participates in interaction among scientist, manager, stakeholder, program staff and consumer.

III. OBJECTIVES AND SCOPE OF SMART GRID DEMAND MONITORING MODEL

We are preparing to design a Demand Monitoring Model. This model is based on distribution consumption and monitoring which intelligently manage the increasing demand for power and optimizing power consumption within the city by having a proper monitor on it. These contribute to scalable, secure software architecture for demand. It will perform the following tasks:

- First and foremost task is to monitor the demand state at the particular area.
- Secondly if the demand exceed then the supply ordered will have some term n condition.
- This will also help in the less consumption of the power.
- Identifies the process of power distribution and demand of power for consumption.
- A proper communication and management standard for demand at a particular area.

The main work done by this model is:

- Analysis
- Decision making

Analysis is the process to study the consumption that occurred daily/monthly by the single consumer, group of consumers and whole consumer in the area. This analysis is based on the data produced daily/monthly and mainly done due to the increasing demand of power. The analysis is based on time and consumption in the area and to get the load status at various level of time, the condition of time can also be observed according to season.

Decision making is done after the whole analysis on the demand areas, priorities/mandatory areas, normal/regular consumption areas on the time basis and other areas. It will confirmed/declined the request of power demand after calculate the following terms:

- Mandatory area on the top priority.
- Time based consumption like industries.

IV. APPLICATION OF SMART GRID DEMAND MONITORING MODEL

This research is done to design a model that makes a smooth linking between network management, demand management, presentation and protection. This will also help in achieving concrete feature (better performance at power consumption, facility of consumer demand and better performance of distributor).

- Distributor

Distributor is one who has the state to supply power to the proper areas using various sources, and try to set the standard of power distribution. At this stage, the distributor will distribute the amount of power demanded by the consumer. But in demand monitoring model the distributor will monitor and set power amount for particular area. This will called as Smart Grid Demand Monitoring Model.

- Consumer

Consumer is one who is going to consume the power, supplied by the distributor on demand and will try to avoid the power wastage and minimize the consumption. The terms and conditions about the limited/ over limited consumption and up gradation the load are set and described by the distributor.

- Consumer contribution- side by side consumer alertness and consumer involvement in Smart Grid tool setting.
- Enhance strength and Active Proficiency- attaining active proficiency by getting enhances strength due to the real-time monitoring of load demand.
- Reaction for Disruption- reaction for an organization power cut and disruption get quick heal with Smart technology.

At last model will furnish a framework for determining tactical intention and application proposal that pillar grid revolution.

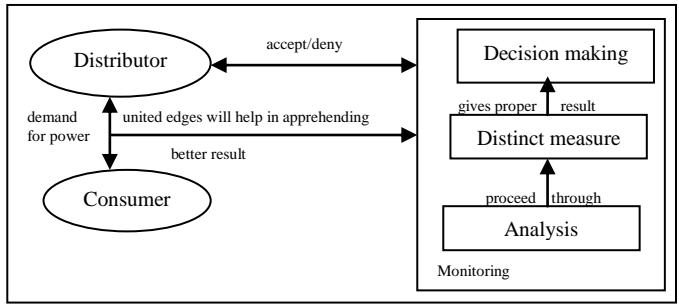

Figure 1.   Proposed Demand Monitoring Model

## V.    CONCLUSION

A model will be structured that stipulate a complete monitoring of demand after analysis and calculating the consumption of different areas. The features of this model will as follows:
- Power superiority- power superiority in relation to its accessibility, voltage solidity, flexibility, resolving the problem from trouble and failure.